\documentclass[useAMS,usenatbib]{mn2e}
\usepackage{txfonts}
\usepackage{graphicx}
\usepackage{natbib}
\usepackage[all]{xy}

\def\del#1{{}}

\newcommand{\ltsima}{$\; \buildrel < \over \sim \;$}
\newcommand{\lsim}{\lower.5ex\hbox{\ltsima}}
\newcommand{\gtsima}{$\; \buildrel > \over \sim \;$}
\newcommand{\gsim}{\lower.5ex\hbox{\gtsima}}

\newcommand{\bra}{\langle}
\newcommand{\ket}{\rangle}

\newcommand{\cll}{{\!\!\!\!\!\!\!}}
\newcommand{\clm}{{\!\!\!\!\!}}

\newcommand{\dd}{\mathrm{d}}

\title[Gravitomagnetic weak lensing and the integrated Sachs-Wolfe effect]
{Weak lensing in the second post-Newtonian approximation:\\ Gravitomagnetic potentials and the integrated Sachs-Wolfe effect}

\author[B. M. Sch\"afer and M. Bartelmann]
{B. M. Sch\"afer$^{1}$\thanks{e-mail: spirou@mpa-garching.mpg.de (BMS); mbartelmann@ita.uni-heidelberg.de (MB)}
and M. Bartelmann$^{2}$\footnotemark[1]\\
$^1$Max-Planck-Institut f\"ur Astrophysik, Karl-Schwarzschild-Stra{\ss}e 1, Postfach 1317, 85741 Garching, Germany\\
$^2$Institut f\"ur theoretische Astrophysik, Universit\"at Heidelberg, Albert-Ueberle-Stra{\ss}e 2, 69120 Heidelberg, Germany}

\begin{document}
\pagerange{\pageref{firstpage}--\pageref{lastpage}}
\pubyear{2003}
\maketitle
\label{firstpage}

% --- abstract --- %
\begin{abstract}
Dark matter currents in the large-scale structure give rise to gravitomagnetic terms in the metric, which affect the light 
propagation. Corrections to the weak lensing power spectrum due to these gravitomagnetic potentials are evaluated by perturbation 
theory. A connection between gravitomagnetic lensing and the integrated Sachs-Wolfe (iSW) effect is drawn, which can be described 
by a line-of-sight integration over the divergence of the gravitomagnetic vector potential. This allows the power spectrum of the 
iSW-effect to be derived within the framework of the same formalism as derived for gravitomagnetic lensing and reduces the 
iSW-effect to a second order lensing phenomenon. The three-dimensional power spectra are projected by means of a generalised 
Limber-equation to yield the angular power spectra. While gravitomagnetic corrections to the weak lensing spectrum are negligible 
at observationally accessible scales, the angular power spectrum of the iSW-effect should be detectable as a correction to the CMB 
spectrum up to multipoles of $\ell\simeq100$ with the {\em Planck}-satellite.
\end{abstract}

% --- keywords --- %

% ############################################################################################## %
% ------------------------------ section: introduction ----------------------------------------- %
\section{Introduction}\label{sect_intro}
Cosmological weak lensing \citep{2001PhR...340..291B} has evolved to be a valuable tool in cosmology. Weak lensing surveys have 
contributed significantly to the determination of the dark matter power spectrum and to the estimation of its amplitude 
$\sigma_8$ \citep{1998MNRAS.296..873S, 2000A&A...358...30V} by the measurement of cosmic shear and have enabled the 
reconstruction of the dark matter distribution in rich clusters of galaxies \cite[e.g. ][]{1993ApJ...404..441K, 
1996A&A...314..707S, 2003mecg.conf..537M}.

So far, only static matter distributions have been considered but from the solution to Maxwell's equations in the framework of 
general relativity it follows that gravitomagnetic potentials generated by moving masses should alter the predictions for light 
deflection \citep{1992grle.book.....S}. While gravitomagnetic corrections to lensing are small, being of order $\upsilon/c$, 
where $\upsilon$ is the velocity of the deflecting mass, they may contribute to the weak cosmological lensing: The cluster 
peculiar velocities following from a cosmological $N$-body simulation like the Hubble-volume simulation 
\citep{2000MNRAS.313..229C, 2001MNRAS.321..372J} are well described by a Gaussian distribution with zero mean and a standard 
deviation of $\sigma_\upsilon \simeq 300~\mathrm{km}/\mathrm{s}$, which is a fraction of $10^{-3}$ of the speed of light. 
Thus, relativistic effects influence the lensing signal appreciably in $\lsim 1$\% of all clusters. In filaments 
\citep{2004astro.ph..6665C} where matter is funneled towards the clusters, velocities are even higher: Infall velocities up 
to a few $10^3~\mathrm{km}/\mathrm{s}$ have been measured.

The integrated Sachs-Wolfe (iSW) effect, or Rees-Sciama (RS) effect \citep{1967ApJ...147...73S, rees_sciama_orig} arises if CMB 
photons encounter time-varying gravitational potentials on their passage from the last-scattering surface to the observer. When 
transversing time-varying potentials, the energy gains and losses a CMB photon experiences in entering and leaving potential 
wells do not cancel exactly. In this way, one expects a net blueshift of CMB photons in forming voids and a net redshift in 
matter-accreting clusters of galaxies. 

The iSW/RS effect has been studied theoretically in individual objects \citep{1990ApJ...355L...5M} and can be used for the 
investigation of cluster mergers \citep{2004A&A...419..439R}. More importantly, it is sensitive to mapping the large-scale 
structure as it highlights the sites of active structure formation \citep{1982MNRAS.198.1033K, 1990MNRAS.247..473M, 
1992PhRvD..46.4193M, 1994ApJ...436....1M, 1996ApJ...460..549S}. Furthermore, the iSW-effect may turn out to be a powerful probe 
for dark energy's influence on structure formation \citep{1996PhRvL..76..575C}, when combined with other tracers of structure. A 
numerical approach has been undertaken by \citet{1995NYASA.759..692T, 1995ApJ...445L..73T}, who followed photons through a 
cosmological $n$-body simulation and carried out the line-of-sight integration numerically. 

The aim of this paper is to determine the corrections to the power spectra of weak lensing quantities caused by gravitomagnetic 
terms and to derive the iSW power spectrum, both by applying perturbation theory. In comparison to preceeding treatments by 
\citet{1996ApJ...460..549S} and \citet{2002PhRvD..65h3518C}, the novel approach taken to determine the iSW power spectrum is by 
relating it to the gravitomagnetic terms in considered in lensing. Gravitomagnetic corrections to lensing have indeed been 
observed by \citet{2003ApJ...598..704F} in imaging radio waves from a quasar on Jupiter, which is an outstanding archivement in 
VLBI astrometry. Gravitomagnetic corrections to lensing in the large-scale structure would only be detectable by their 
$n$-point statistics or by topological measures like Minkowski functionals, that would be especially sensitive to the effect's 
intrinsic non-Gaussianity. Concerning the iSW-effect, there are a quite a few reports on its detection in WMAP data in cross 
correlation with various populations of tracer objects \citep{2004PhRvD..69h3524A, 2003ApJ...597L..89F, 2004Natur.427...45B, 
2004ApJ...608...10N, 2004astro.ph..6004H}, but so far it has not been possible to derive values for single multipoles based on CMB 
data alone.

The paper is structured as follows: After a compilation of key formulae and the derivation of Limber's equation for vector fields 
in Sect.~\ref{sect_key_formulae}, the power spectrum of weak gravitational lensing is considered and the correction terms due to 
gravitomagnetic potentials are worked out by perturbation theory in Sect.~\ref{sect_gravitomagnetic}. Then, the iSW-effect is 
related to gravitomagnetic lensing and its power spectrum is subsequently derived in a perturbative approach in 
Sect.~\ref{sect_rees_sciama}. The results are summarised in Sect.~\ref{sect_summary}.

% ############################################################################################## %
% ------------------------------ section: gravitomagnetic lensing -----------------------------  %
\section{Key formulae}\label{sect_key_formulae}
The assumed cosmological model is the standard \mbox{$\Lambda$CDM} cosmology, which has recently been supported by 
observations of the WMAP satellite\footnote{\tt http://map.gsfc.nasa.gov/} \citep{2003astro.ph..2209S}. Parameter values have 
been chosen as $\Omega_\mathrm{M} = 0.3$, $\Omega_\Lambda =0.7$, $H_0 = 100\,h\,\mbox{km~}\mbox{s}^{-1}\mbox{ Mpc}^{-1}$ with 
$h = 0.7$, $\Omega_\mathrm{B} = 0.04$, $n_\mathrm{s} =1$ and $\sigma_8=0.9$.

% ============================================================================================== %
\subsection{Structure formation}
The cosmic density field $\rho$ given in terms of the dimensionless density perturbation $\delta = (\rho-\bra\rho\ket) / 
\bra\rho\ket$, where $\bra\rho\ket$ is the average density of matter. The 2-point correlation properties of the overdensity 
field $\delta$ are described by the power spectrum $P(k)$:
\begin{equation}
\bra\delta(\bmath{k})\delta^*(\bmath{k^\prime})\ket=(2\pi)^3\delta_D(\bmath{k}-\bmath{k^\prime})P(k)\mbox{, where}
\end{equation}
\begin{equation}
\delta({\bmath{k}}) = \int\dd^3 x\:\delta({\bmath{x}}) \exp(i \bmath{k}\bmath{x})
\end{equation}
is the Fourier transform of the overdensity field $\delta$. The normalisation of the power spectrum $P(k)$ is given by the 
parameter $\sigma_8$, i.e. the variance of $\delta$ on scales of $R=8~\mathrm{Mpc}/h$:
\begin{equation}
\sigma_R^2 = \frac{1}{2\pi^2}\int_0^\infty\dd k\:k^2 W^2(kR) P(k)\mbox{.}
\end{equation}
Here, $W$ is a window function of top-hat shape, the Fourier-transform of which is given by:
\begin{equation}
W(x) = \frac{3}{x^3}\left[\sin(x)-x\cos(x)\right] = \frac{3}{x}J_1(x)\mbox{.}
\end{equation}
The shape of the power spectrum $P(k) \propto k^{n_\mathrm{s}}\cdot T^2(k)$ is well approximated by the transfer functions 
$T(k)$ suggested by \citet{1986ApJ...304...15B}. They read in case of adiabatic initial conditions:
\begin{equation}
T(q) = \frac{\ln(1+2.34q)}{2.34q}\left[1+3.89q+(16.1q)^2+(5.46q)^3+(6.71q)^4\right]^{-\frac{1}{4}}
\end{equation}
The wave vector $k$ is commonly divided by the shape parameter $\Gamma$ introduced by \citet{1992MNRAS.258P...1E} for CDM 
models and extended to models with $\Omega\not=1$ by \citet{1995ApJS..100..281S}:
\begin{equation}
q = q(k) = \frac{k/\mathrm{Mpc}^{-1}h}{\Gamma}\mbox{ with }
\Gamma=\Omega_M h \exp\left(-\Omega_B\cdot\left[1+\frac{\sqrt{2 h}}{\Omega_M}\right]\right)\mbox{.}
\end{equation}
In linear structure formation, each Fourier-mode grows independently and at the same rate. The time dependence of the 
overdensity field $\delta$ can be described by the growth function $D(a)$:
\begin{equation}
\delta(a)=\delta_0 D(a)\mbox{ with }D(a) = a\frac{d^\prime(a)}{d^\prime(1)}\mbox{.}
\label{eqn_dplus_def}
\end{equation}
The shape of $d^\prime(a)$ is well approximated by the formula suggested by \citet{1992ARA&A..30..499C}:
\begin{equation}
d^\prime(a) = \frac{5}{2}\Omega_M(a)\left[\Omega_M^{4/7}(a)-\Omega_\Lambda(a)+
\left(1+\frac{\Omega_M(a)}{2}\right)\left(1+\frac{\Omega_\Lambda(a)}{70}\right)\right]^{-1}\mbox{.}
\end{equation}

% ============================================================================================== %
\subsection{Dark matter currents}
The continuity equation $\dot{\rho}=-\mathrm{div}(\rho\bmath{\upsilon})$ requires the existence of large-scale coherent matter 
streams $\bmath{j}=\rho\bmath{\upsilon}$ superimposed on the Hubble flow due to the formation of structure. In Fourier space, 
the relation between density and velocity reads in the Eulerian frame in linear approximation:
\begin{equation}
\bmath{\upsilon}(\bmath{k}) 
= -i a H(a) f(\Omega) \frac{\bmath{k}}{k^2}\delta(\bmath{k}) 
= -i \dot{a} f(\Omega) \frac{\bmath{k}}{k^2}\delta(\bmath{k})\mbox{.}
\end{equation}
The $1/k$-dependence causes cosmological velocities to come predominantly from perturbations on larger scales in comparison to 
those that dominate the density field. $H(a)=\dd\ln(a)/\dd t$ is Hubble's function. The function $f$ describes the dependence 
of the equation of continuity on cosmic time and is a function of the mass density $\Omega$ \citep{1980lssu.book.....P, 
1991MNRAS.251..128L}:
\begin{equation}
f(\Omega) = \frac{\dd\ln\delta}{\dd\ln a} = \frac{\dd\ln D(a)}{\dd\ln a} \simeq\Omega(a)^{0.6}
\end{equation}
In analogy to eqn.~(\ref{eqn_dplus_def}), time evolution of of dark matter current velocities in the comoving frame is 
described by $G(a)$,
\begin{equation}
G(a) = \frac{g^\prime(a)}{g^\prime(1)}\mbox{ with } g^\prime(a) \equiv H(a) f(\Omega)\mbox{.}
\end{equation}
The theory of peculiar velocity fields is reviewed in detail in \citet{1994ARA&A..32..371D} and \citet{1995PhR...261..271S}.

In general, the effects considered here are sensitive to density weigted velocities. The Fourier transform of vector fields 
$\bmath{q(\bmath{x})} = \delta(\bmath{x})\bmath{\upsilon}(\bmath{x})$ can be derived with the convolution theorem:
\begin{eqnarray}
\bmath{q}(\bmath{k}) 
& = & \int\dd^3 x\:\bmath{q}(\bmath{x})\exp(i\bmath{k}\bmath{x}) \\
& = & \frac{1}{2}\int\frac{d^3 p}{(2\pi)^3}
\left[
\bmath{\upsilon}(\bmath{p})\delta(\bmath{k}-\bmath{p}) + \bmath{\upsilon}(\bmath{k}-\bmath{p})\delta(\bmath{p})
\right]\mbox{,}
\label{eqn_dv_fft}
\end{eqnarray}
where the integrand has been symmetrised in eqn.~(\ref{eqn_dv_fft}).

% ============================================================================================== %
\subsection{Limber's equation for vector fields}
For the derivation of the angular power spectrum of the gravitomagnetic corrections to weak cosmological lensing or that of the 
iSW-effect, a variant of Limber's equation is necessary that is able to deal with projections of vector fields 
$\bmath{q}(\bmath{x})$ instead of scalar fields. The derivation presented here is generalised from \citet{1987ApJ...322..597V}. 
Consider a vector field $\bmath{q}(\bmath{x})$ and its Fourier transform $\bmath{q}(\bmath{k})$:
\begin{equation}
\bmath{q}(\bmath{x}) = \int\frac{\dd^3 k}{(2\pi)^3}\:\bmath{q}(\bmath{k})\exp(-i\bmath{kx})
\end{equation}
Any effect $\kappa$ in question is assumed to measure a projection of $\bmath{q}(\bmath{x})$ on the line-of-sight, 
where $\bmath{e}$ is a unit tangent vector on the photon geodesic. $W(w)$ is a general weighing function dependent on the 
comoving distance $w$ which describes its redshift dependence and is later to be replaced by e.g. the lensing efficiency 
function:
\begin{equation}
\kappa 
= \int_0^{w_\mathrm{max}}\cll\dd w\: W(w) \left[\bmath{e}\cdot\bmath{q}\right]
= \int_0^{w_\mathrm{max}}\cll\dd w\: W(w) \left[\bmath{e}\cdot\bmath{q}(\bmath{k})\right]
\int\frac{\dd^3 k}{(2\pi)^3}\:\exp(-i\bmath{kx})
\end{equation}
The decomposition of the projected field $\kappa(\bmath{\theta})$ into spherical harmonics $Y_{\ell m}(\bmath{\theta})$ is:
\begin{equation}
\kappa(\bmath{\theta}) = \sum_{\ell=0}^{\infty}\sum_{m=-\ell}^{+\ell} \kappa_{\ell m} Y_{\ell m}(\bmath{\theta})
\leftrightarrow
\kappa_{\ell m} = \int_{4\pi}\!\!\dd\Omega\: \kappa(\bmath{\theta}) Y_{\ell m}^*(\bmath{\theta})\mbox{ with}
\end{equation}
\begin{equation}
Y_{\ell m}(\bmath{\theta}) = 
\sqrt{\frac{2\ell + 1}{4\pi}} \sqrt{\frac{(\ell - \left|m\right|)!}{(\ell + \left|m\right|)!}} 
P_{\ell m}\left(\cos\theta\right) \exp\left(i m \phi\right)\mbox{.}
\end{equation}
In the random phase approximation, one obtains for the variance $\bra\left|\kappa_{\ell m}\right|^2\ket$ of 
$\kappa(\bmath{\theta})$ in two directions $\bmath{e}_1$ and $\bmath{e}_2$:
\begin{eqnarray}
\bra\left|\kappa_{\ell m}\right|^2\ket & = & 
\int_0^{w_\mathrm{max}}\cll\dd w_1\: W(w_1)
\int_{4\pi}\dd\Omega_1\:Y_{\ell m}(\bmath{e}_1) \\
& & \int_0^{w_\mathrm{max}}\cll\dd w_2\: W(w_2)
\int_{4\pi}\dd\Omega_2\:Y_{\ell m}^*(\bmath{e}_2) \nonumber\\
& & \int\frac{\dd^3 k}{(2\pi)^3}\: \exp(-i\bmath{k}\bmath{e}_1 w_1) \exp(i\bmath{k}\bmath{e}_2 w_2)
\bra\left[\bmath{e}_1\bmath{q}(\bmath{k})\right]\left[\bmath{e}_2\bmath{q}^*(\bmath{k})\right]\ket
\label{eqn_spectrum_vector}
\nonumber\mbox{.}
\end{eqnarray}
According to the cosmological principle, there is no preferred orientation, which allows to replace 
$\bra\left|a_{\ell m}\right|^2\ket$ with its average value over all $m$ for a given $\ell$:
\begin{equation}
C_\kappa(\ell) = \frac{1}{2\ell+1}\sum_{m=-\ell}^{+\ell}\bra\left|\kappa_{\ell m}\right|^2\ket\mbox{.}
\end{equation}
The vector field $\bmath{q}(\bmath{k})$ can be separated into components parallel and perpendicular to the line-of-sight 
$\bmath{e}$:
\begin{equation}
\bmath{q} = \bmath{q}_\parallel + \bmath{q}_\perp
\mbox{ with }
\bmath{q}_\parallel = \bmath{e}(\bmath{q}\cdot\bmath{e})
\mbox{ and }
\bmath{q}_\perp = \bmath{q}-\bmath{q}_\parallel = \bmath{e}\times(\bmath{q}\times\bmath{e})
\mbox{.}
\end{equation}
For the projections $\bmath{e}\cdot\bmath{q}_\perp = 0$ and $\bmath{e}\times\bmath{q}_\parallel = 0$ are valid. 
Eqn.~(\ref{eqn_spectrum_vector}) is further simplified by choosing the coordinate system in a way that the 
$z$-coordinate is parallel to the wave vector, $\bmath{e}_\mathrm{z}\parallel\bmath{k}$. Introducing spherical coordinates 
$(\theta,\phi)$ and putting $x=\cos\theta$ on obtains:
\begin{equation}
\bmath{q}_\parallel\cdot\frac{\bmath{k}}{k} = x q_\parallel \mbox{ and }
\bmath{q}_\perp\cdot\frac{\bmath{k}}{k} = \sqrt{1-x^2}\exp(-i\phi) q_\perp
\end{equation}
Furthermore, with $\exp(i\bmath{k}\bmath{e} w) = \exp(ikxw)$, the expression for the correlator is separated into:
\begin{displaymath}
\bra\bmath{q}(\bmath{k})\cdot\bmath{q}^*(\bmath{k})\ket = 
\end{displaymath}
\begin{equation}
x_1 x_2 \bra q_\parallel(\bmath{k})\cdot q^*_\parallel(\bmath{k})\ket + 
\sqrt{1-x_1^2} e^{-i\phi_1}\sqrt{1-x_2^2} e^{i\phi_2} \bra q_\perp(\bmath{k})q^*_\perp(\bmath{k})\ket
\mbox{.}
\end{equation}
With these simplifications, the integrals over the azimuthal angles $\phi_1$ and $\phi_2$ in eqn.~(\ref{eqn_spectrum_vector}) can be carried out. Inserting the orthonormality relation $\int_0^{2\pi}\dd\phi\exp\left[i(n-m)\phi\right]=2\pi\delta_{mn}$ reduces the summation 
over $m$ to a single term, which is $m=0$ for the components parallel to the line-of-sight and $\left|m\right|=1$ for the 
components perpendicular to the line-of-sight. The final expression for the power spectrum $C_\kappa(\ell)$ is now split into the two orthogonal projections:
\begin{equation}
C_\kappa(\ell) = C_\kappa^\parallel(\ell) + C_\kappa^\perp(\ell)\mbox{.}
\end{equation}

% ---------------------------------------------------------------------------------------------- %
\subsubsection{Components parallel to the line-of-sight $C_\kappa^\parallel(\ell)$}
For the power spectrum $C_\kappa^\parallel(\ell)$ of the components of $\bmath{q}_\parallel$ parallel to the line-of-sight, one 
obtains:
\begin{eqnarray}
C_\kappa^\parallel(\ell) & = &
\frac{2}{(2\pi)^2}\int\dd k\:k^2 
\int_0^{w_\mathrm{max}}\cll\dd w_1 W(w_1)\int_0^{w_\mathrm{max}}\cll\dd w_2 W(w_2) \\
& & \int_{-1}^{+1}\clm\dd x_1 \exp(-i k x_1 w_1) \int_{+1}^{+1}\clm\dd x_2 \exp(i k x_2 w_2) \nonumber\\
& & \left[x_1 P_{\ell 0}(x_1) x_2 P_{\ell 0}(x_2)\right]
\bra q_\parallel(\bmath{k},w_1)q_\parallel^*(\bmath{k},w_2)\ket\mbox{.}\nonumber
\end{eqnarray}
The $\dd x_1$- and $\dd x_2$-integrations can be performed by taking advantage of the connection between Bessel functions and 
Legendre polynomials \citep{watson_bessel, 1972hmf..book.....A}:
\begin{equation}
J_\ell(z) = \frac{1}{2i^\ell}\int_{-1}^{+1}\clm\dd x\: P_\ell(x) \exp(i z x)\mbox{,}
\label{eqn_bessel_legendre}
\end{equation}
which can be can be generalised to give:
\begin{equation}
\int_{-1}^{+1}\clm\dd x\: x^n P_\ell(x)\exp(i z x) = \frac{1}{i^n}\frac{\dd^n}{\dd z^n} J_\ell(z)\mbox{.}
\label{eqn_bessel_legendre_recurrence}
\end{equation}
Inserting formula~(\ref{eqn_bessel_legendre_recurrence}) for $n=1$ yields the final result:
\begin{eqnarray}
C_\kappa^\parallel(\ell) & = & 
\frac{1}{(2\pi)^2}\int\dd k
\int_0^{w_\mathrm{max}}\cll\dd w_1 W(w_1)\int_0^{w_\mathrm{max}}\cll\dd w_2 W(w_2) \nonumber\\
& & \left[\frac{\dd}{\dd w_1} J_\ell(k w_1)\right]\left[\frac{\dd}{\dd w_2} J_\ell(k w_2)\right]
\bra q_\parallel(\bmath{k},w_1) q_\parallel^*(\bmath{k},w_2)\ket\mbox{.}
\label{eqn_proj_parallel}
\end{eqnarray}

% ---------------------------------------------------------------------------------------------- %
\subsubsection{Components perpendicular to the line-of-sight $C_\kappa^\perp(\ell)$}
After reducing the summation to $\left|m\right|=1$, the power spectrum $C_\kappa^\perp(\ell)$ of the components of 
$\bmath{q}_\perp$ perpendicular to the line-of-sight reads:
\begin{eqnarray}
C_\kappa^\perp(\ell) & = & 
\frac{1}{(2\pi)^2\ell(\ell+1)}\int\dd k\:k^2 
\int_0^{w_\mathrm{max}}\cll\dd w_1 W(w_1)\int_0^{w_\mathrm{max}}\cll\dd w_2 W(w_2)  \nonumber\\
& & \int_{-1}^{+1}\clm\dd x_1 \exp(-i k x_1 w_1)
\int_{+1}^{+1}\clm\dd x_2 \exp(i k x_2 w_2) \nonumber\\
& & \left[\sqrt{1-x_1^2} P_{\ell 1}(x_1) \sqrt{1-x_2^2} P_{\ell 1}(x_2)\right]
\bra q_\perp(\bmath{k},w_1) q_\perp^*(\bmath{k},w_2)\ket\mbox{.}\nonumber
\end{eqnarray}
The integration over the polar angles $x_1$ and $x_2$ is slightly more complicated than the previous case. Inserting the
definition of the associated Legendre polynomials $P_{\ell m}$ for $m = 1$ gives another factor of $\sqrt{1-x^2}$:
\begin{equation}
P_{\ell m}(x) = (-1)^m (1-x^2)^\frac{m}{2}\frac{\dd^m P_\ell(x)}{\dd x^m} \rightarrow
P_{\ell 1}(x) = -\sqrt{1-x^2}\frac{\dd P_\ell(x)}{\dd x}\mbox{.}
\end{equation}
The derivative of the Legendre polynomial can be replaced via
\begin{equation}
(1-x^2)\frac{\dd}{\dd x} P_\ell(x) = \ell\left[P_{\ell-1}(x) - x P_\ell(x)\right]\mbox{,}
\end{equation}
and the integration be carried out by inserting relation~(\ref{eqn_bessel_legendre_recurrence}). Then, the two Bessel 
functions can be combined by using the Bessel function's derivative relation:
\begin{equation}
\frac{\dd}{\dd z}\left[z^\ell J_\ell(z)\right] = z^\ell J_{\ell-1}(z) \rightarrow 
\frac{J_\ell(z)}{z} = \frac{1}{\ell}\left[J_{\ell-1}-\frac{\dd}{\dd z}J_\ell(z)\right]\mbox{,}
\end{equation}
which yields the formula:
\begin{equation}
\int_{-1}^{+1}\clm\dd x\:\sqrt{1-x^2} P_{\ell 1}(x)\exp(i z x) = \ell(\ell+1)\frac{J_\ell(z)}{z}\mbox{.}
\end{equation}
This relation allows the final result to be written as:
\begin{eqnarray}
C_\kappa^\perp(\ell) & = & 
\frac{\ell(\ell+1)}{(2\pi)^2}\int\dd k 
\int_0^{w_\mathrm{max}}\cll\dd w_1 W(w_1)\int_0^{w_\mathrm{max}}\cll\dd w_2 W(w_2) \nonumber\\
& & \left[\frac{J_\ell(k w_1)}{w_1} \frac{J_\ell(k w_2)}{w_2}\right]
\bra q_\perp(\bmath{k},w_1) q_\perp^*(\bmath{k},w_2)\ket\mbox{.}
\label{eqn_proj_perp}
\end{eqnarray}

% ############################################################################################## %
% ------------------------------ section: gravitomagnetic lensing -----------------------------  %
\section{Gravitomagnetic lensing}\label{sect_gravitomagnetic}

% ============================================================================================== %
\subsection{Definitions}\label{def_gravitomagnetic}
Light propagation through a slowly moving perfect fluid can be described by an effective refractive index $n_\mathrm{eff}$ 
which follows from the post-Newtonian expansion of the Raychaudhuri-equation to second order for a weakly perturbed space-time \citep{1992grle.book.....S}:
\begin{equation}
n_\mathrm{eff} = 1 - \frac{2}{c^2}\Phi + \frac{4}{c^3}\bmath{A}\cdot\bmath{e}\mbox{.}
\label{eqn_def_neff}
\end{equation}
Here, $\Phi$ is the scalar potential and $\bmath{A}$ are the gravitomagnetic vector potentials. $\bmath{e}$ denotes a unit 
tangent vector along the photon geodesic. In this approximation, the metric takes account of the matter density $\rho$ and the 
matter current densities $\bmath{j} = \rho\bmath{\upsilon}$ (i.e. terms of order $\upsilon/c$), but neglects the stresses 
$T_{ij}=\rho\upsilon_i\upsilon_j+p\delta_{ij}$. The smallness of these terms (being of order $\upsilon^2/c^2$) makes them 
unobservable, but they would be sensitive to the velocity tensor $\upsilon_i\upsilon_j$ , i.e. to shear flows, velocity 
dispersions and turbulence.

In the near zone of a system of slowly moving bodies the retardation can be neglected; in this case the expressions for $\Phi$ 
and $\bmath{A}$ are given as solutions to Laplace's equation:
\begin{eqnarray}
\Delta\Phi(\bmath{r}) = 4\pi G\rho(\bmath{r}) & \leftrightarrow & 
\Phi(\bmath{r}) = -G\int\dd^3r^\prime\frac{\rho(\bmath{r}^\prime)}{\left|\bmath{r}-\bmath{r}^\prime\right|} \\
\Delta\bmath{A}(\bmath{r}) = 4\pi G\bmath{j}(\bmath{r}) & \leftrightarrow &
\bmath{A}(\bmath{r}) = -G\int\dd^3r^\prime\frac{\bmath{j}(\bmath{r}^\prime)}{\left|\bmath{r}-\bmath{r}^\prime\right|}\mbox{.}
\label{eqn_lensing_defa}
\end{eqnarray}
The dark matter flux $\bmath{j}$ is defined as the momentum density $\bmath{j}\equiv\rho\bmath{\upsilon}$.

An expression for $\dd \bmath{e}/\dd w$, i.e. the change in propagation direction, follows from the variational principle 
$\delta\int\dd s\, n_\mathrm{eff} = 0$. $s$ denotes an affine parameter. The deflection angle $\bmath{\alpha}$, being defined 
as $\bmath{\alpha}=\bmath{e}_\mathrm{initial}-\bmath{e}_\mathrm{final}$ can be obtained by integration:
\begin{equation}
\bmath{\alpha} = 
\frac{2}{c^2}\int\dd s\: \nabla_\perp\Phi - 
\frac{4}{c^3}\int\dd s\: \bmath{e}\times\mathrm{rot}\bmath{A}\mbox{.}
\label{eqn_lensing_alpha}
\end{equation}
The derivative perpendicular to the line-of-sight is defined via 
$\nabla_\perp\Phi \equiv \nabla\Phi - \bmath{e}(\bmath{e}\cdot\nabla\Phi)$. The first contribution to $\bmath{\alpha}$ in 
eqn.~(\ref{eqn_lensing_alpha}) corresponds to the attraction $\bmath{g}$ towards the deflecting mass via $\bmath{g}=-\nabla\Phi$. 

The second term, however, is due to the gravitomagnetic fields generated by the matter current densities $\bmath{j}$. This 
contribution is related to the dragging of inertial frames which gives rise to the precession of orbiting spinning tops in the 
particular case of rotation of the field-generating body (Lense-Thirring precession, to be measured by Gravity
Probe B\footnote{\tt http://www.gravityprobeb.com}). This formalism has been applied to various astrophysical systems, namely by 
\citet{1983A&A...124..175I} to gravitational light deflection of a rotating galaxy and by \citet{2003MNRAS.344..942S}, who 
considered light deflection on rotating stars. Furthermore, corrections to the deflection angle in galactic microlensing due to 
moving lenses have been evaluated by \citet{2004astro.ph.10183H}.

% ============================================================================================== %
\subsection{Gravitomagnetic lensing by the large-scale structure}\label{gravitomagnetic_lss}
Adopting the Born-approximation, which states that the gravitational light deflection is weak such that the integral in 
eqn.~(\ref{eqn_lensing_alpha}) can be evaluated along a straight line instead of the photon geodesic itself, it can be seen 
that gravitational lensing is insensitive to derivatives of the potentials along the line-of-sight. Working out the deflection angles $\bmath{\alpha}$ and the tidal matrix $\psi_{ij}=\partial\alpha_i/\partial x_j$ while neglecting derivatives along the line-of-sight yields formulae analogous to the case of static lensing, but with the gravitational potential $\Phi$ replaced by $\Phi-\frac{2}{c} A_\parallel$. Thus, the sources of gravitational light deflection are the matter distribution $\delta$ and the component of the matter flux $j_\parallel$ parallel to the line-of-sight. The gravitational light deflection is stronger, if an object is moving towards the observer, because the photon stays in the interaction potential for a longer period of time, and vice versa. 

\begin{figure}
\resizebox{\hsize}{!}{\includegraphics{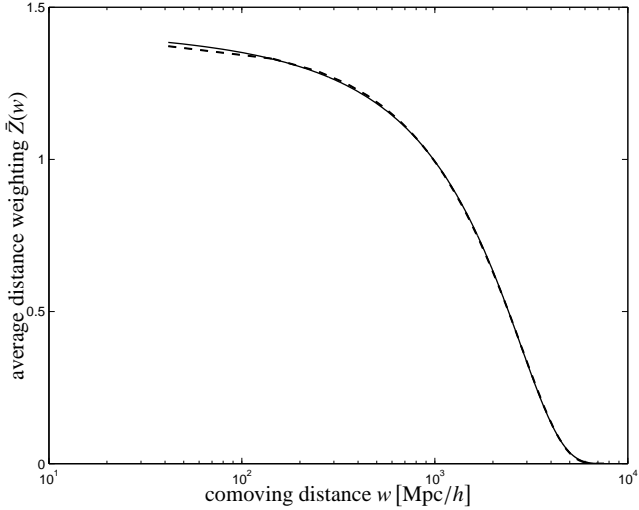}}
\caption{The redshift weighting function $\bar{Z}(w)$ (c.f. eqn.~(\ref{eqn_zsource}), rendered as a dashed line), and the 
fitting formula~(\ref{eqn_zsource_fit}) (solid line) in comparison.}
\label{fig_zsource}
\end{figure}

With the source term $\delta + \frac{2}{c}j_\parallel$, one obtains for the lensing convergence $\kappa$ up to the comoving 
distance $w$ \citep{2001PhR...340..291B}:
\begin{equation}
\kappa(\bmath{\theta},w)=\frac{3H_0^2\Omega_0}{2c^2}\int_0^{w} \dd w^\prime
\frac{f_K(w^\prime) f_K(w_\mathrm{max} - w^\prime)}{f_K(w_\mathrm{max}) a(w^\prime)}\left(\delta+\frac{2}{c}j_\parallel\right)
\mbox{.}
\end{equation}
where $f_K(w) = w$, if spatial hypersurfaces are flat, which is the case for $\Omega_M+\Omega_\Lambda=1$. The redshift 
distribution of lensed population of background sources such as faint blue galaxies is described by the distribution $p(z)\dd z$, 
being recast in comoving distance, $Z(w)\dd w = p(z)\dd z$. Then, the average influence $\bar{Z}(w)$ of the lever arms of the 
optical path for a given configuration of source and lens is given by:
\begin{equation}
\bar{Z}(w) = \int_w^{w_\mathrm{max}}\cll\dd w^\prime\: Z(w^\prime)\frac{f_K(w^\prime-w)}{f_K(w^\prime)}\mbox{.}
\label{eqn_zsource}
\end{equation}
In this work, we assume the generic distribution in redshift $z$ for faint blue galaxies \citep[c.f.][]{1997ARA&A..35..389E},
\begin{equation}
p(z)\dd z = p_0 z^2 \exp(-z^\beta)\mbox{ with } \frac{1}{p_0} = \frac{1}{\beta}\Gamma\left(\frac{3}{\beta}\right)
\mbox{.}
\end{equation}
with mean redshift $\bra z\ket=\Gamma(4/\beta)/\Gamma(3/\beta)\simeq1.5$ and most likely redshift 
$z_\mathrm{max}=(2/\beta)^{1/\beta}\simeq1.21$ for $\beta=3/2$. For the average convergence $\bar{\kappa}$, the final result 
reads:
\begin{eqnarray}
\bar{\kappa}(\bmath{\theta}) & = & \int_0^{w_\mathrm{max}}\cll\dd w\:Z(w)\kappa(\theta,w)\nonumber\\
& = & \frac{3 H_0^2\Omega_0}{2c^2}\int_0^{w_\mathrm{max}}\cll \dd w\: 
\bar{Z}(w)\frac{f_K(w)}{a(w)}\left(\delta+\frac{2}{c}j_\parallel\right)\mbox{.}
\label{eqn_kappa_cosmo}
\end{eqnarray}
For $\bar{Z}(w)$, the phenomenological fitting formula
\begin{equation}
\bar{Z}(w) \simeq Z_0 \exp\left(-\frac{1}{1-\left[\log(w/w_0)\right]^b}\right)\mbox{,}
\label{eqn_zsource_fit}
\end{equation}
with $Z_0 = 1.441$, $b = 3.186$ and $w_0=2314~\mathrm{Mpc}/h$ is used, which yields excellent agreement with the properly 
evaluated function, as shown by Fig.~\ref{fig_zsource}. The fitting formula alleviates the need of numerically carrying out 
the integration in eqn.~(\ref{eqn_zsource}) when projecting the dark matter power spectrum.

% ============================================================================================== %
\subsection{Perturbative treatement}
When considering gravitomagnetic corrections to gravitational lensing, the source term $\delta$ of static lensing has to be 
replaced by $q_\parallel = (1+\frac{2}{c}v_\parallel)\delta$. It should be emphasised, that the fluctuations in a weak lensing 
shear field are predominantly caused by modes in $k$-space, that are propagating perpendicularly to the line-of-sight 
\citep{1991MNRAS.251..600B}. Evaluating the correlator $\bra q_\perp(\bmath{k},w_1) q^*_\perp(\bmath{k},w_2)\ket$ yields apart 
from the dominating 2-point term,
\begin{equation}
\bra q_\perp(\bmath{k},w_1) q_\perp(\bmath{k},w_2)\ket_\mathrm{2pt} = 
D(w_1) D(w_2) \bra\delta(\bmath{k})\delta^*(\bmath{k})\ket\mbox{,}
\label{eqn_lensing_two}
\end{equation}
contributions of 3- and 4-point terms. The 2-point term stated in eqn.~(\ref{eqn_lensing_two}) is of order unity and is the basis 
of the conventional theory of static gravitational lensing. In the perturbative treatment, the coupling of $\bmath{k}$-modes in 
nonlinear structure growth is neglected, integrations are implicitly taken to be restricted to quasi-linear scales.

% ---------------------------------------------------------------------------------------------- %
\subsubsection{3-point term}
The 3-point term $\bra q_\perp(\bmath{k},w_1) q^*_\perp(\bmath{k},w_2)\ket_\mathrm{3pt}$ consists of four 
contributions and is of order $\upsilon/c$ compared to the 2-point term (c.f. eqn.~(\ref{eqn_lensing_two})):
\begin{equation}
\bra q_\perp(\bmath{k},w_1) q^*_\perp(\bmath{k},w_2)\ket_\mathrm{3pt} =
\frac{1}{c} \int\frac{\dd^3 p}{(2\pi)^3}\left\{\vphantom{\int}\right.
\label{eqn_lensing_three}
\end{equation}
\begin{displaymath}
\bra \delta(\bmath{-k},w_1) \upsilon_\perp(\bmath{p},w_2) \delta(\bmath{k}-\bmath{p},w_2) \ket +
\bra \delta(\bmath{-k},w_1) \upsilon_\perp(\bmath{k}-\bmath{p},w_2) \delta(\bmath{p},w_2) \ket + 
\end{displaymath}
\begin{displaymath}
\bra \delta(\bmath{k},w_2) \upsilon_\perp(\bmath{-p},w_1) \delta(\bmath{p}-\bmath{k},w_1) \ket +
\bra \delta(\bmath{k},w_2) \upsilon_\perp(\bmath{p}-\bmath{k},w_1) \delta(\bmath{-p},w_1) \ket
\left.\vphantom{\int}\right\}
\end{displaymath}
Here, the relations $\delta^*(\bmath{k}) = \delta(-\bmath{k})$ and $\bmath{\upsilon}^*(\bmath{k}) = 
\bmath{\upsilon}(-\bmath{k})$ were inserted, which hold for real fields. By using this fact, the condition 
$\sum_i\bmath{k}_i=\bmath{0}$ is fulfilled which allows the 3-point correlators in eqn.~(\ref{eqn_lensing_three}) to be 
reduced to products of 2-point correlators by virtue of the formulae derived in Appendix~\ref{sect_appendix}.
This yields four terms of the type $\bra\upsilon_\perp\delta\ket\bra\delta\delta\ket/c$ and two contributions 
$\bra\upsilon_\perp\delta\ket^2/c^2$ of second order. 

The correlation function can then be projected onto a plane perpendicular to wave vector $\bmath{k}$: The component of the 
velocity in the celestial plane is given by $\bmath{\upsilon}_\perp=\bmath{k}\times(\bmath{\upsilon}\times\bmath{k}) / k^2$ and 
hence $\upsilon_\perp = \upsilon\sin\theta = \upsilon\sqrt{1-x^2}$, with $x=\cos\theta$, where $\theta$ is the angle of separation 
between $\bmath{p}$ and $\bmath{k}$. In doing this, the contributions of the type $\bra\upsilon_\perp\delta\ket^2/c^2$ vanish 
because they contain a multiplicative term $\bra\delta(\bmath{k}) \upsilon(\bmath{k})\ket$, which is a vector field collinear to 
$\bmath{k}$. The remaining terms can be combined to give:
\begin{eqnarray}
(2\pi)^3\bra q_\perp(\bmath{k},w_1) q^*_\perp(\bmath{k},w_2)\ket_\mathrm{3pt} = 
\frac{4\pi}{c} D(w_1) D(w_2) \left[g^\prime(w_1) + g^\prime(w_2)\right]
\nonumber
\end{eqnarray}
\begin{eqnarray}
\int p^2\dd p\int_{-1}^{+1}\clm\dd x \sqrt{1-x^2} \left\{\vphantom{\int}\right.
P(\left|\bmath{p}\right|) P(\left|\bmath{p}-\bmath{k}\right|) M(\bmath{p},\bmath{p}-\bmath{k})
\left[\frac{1}{\left|\bmath{p}\right|} + \frac{p}{\left|\bmath{p}-\bmath{k}\right|^2}\right]
\nonumber
\end{eqnarray}
\begin{equation}
+P(\left|\bmath{k}\right|)
\left[M(\bmath{k},-\bmath{p}) \frac{P(\left|\bmath{p}\right|)}{\left|\bmath{p}\right|}  +
M(\bmath{k},\bmath{p}-\bmath{k}) \frac{p}{\left|\bmath{p}-\bmath{k}\right|^2} P(\left|\bmath{p}-\bmath{k}\right|)\right]
\left.\vphantom{\int}\right\}
\label{eqn_threept}
\nonumber
\end{equation}
In the integrand of eqn.~(\ref{eqn_threept}), the replacement $\left|\bmath{p}-\bmath{k}\right|^2 = k^2-2kpx +p^2$ can be 
inserted. Additionally, the time evolution of the velocity-density cross correlation function,
\begin{equation}
\bra\upsilon_\perp(\bmath{k},w_1)\delta^*(\bmath{k},w_2)\ket = 
g^\prime(w_1) D(w_2)\bra\upsilon_\perp(\bmath{k})\delta^*(\bmath{k})\ket
\mbox{,}
\end{equation}
was inserted. The function $M(\bmath{p},\bmath{p}^\prime)$ is defined as:
\begin{equation}
M(\bmath{p},\bmath{p}^\prime) = 
\frac{10}{7} + 
\frac{\bmath{p}\bmath{p}^\prime}{p p^\prime}\left(\frac{p}{p^\prime}+\frac{p^\prime}{p}\right) + 
\frac{4}{7}\left(\frac{\bmath{p}\bmath{p}^\prime}{p p^\prime}\right)^2\mbox{.}
\end{equation}

It should be emphasised, that this 3-point correlator does not take account of the evolution of non-Gaussian features 
in the correlation function $\bra\delta(\bmath{k}_1)\delta(\bmath{k}_2)\delta(\bmath{k}_3)\ket$ and their influence on lensing 
determined by \citet{1997ApJ...484..560J, 1997A&A...324...15B} and \citet{2003MNRAS.340..580T, 2003MNRAS.344..857T}, which 
strongly affects weak lensing quantities on small angular scales.

% ---------------------------------------------------------------------------------------------- %
\subsubsection{4-point term}
The last contribution to the weak lensing power spectrum evoked by gravitomagnetic corrections is the 4-point term 
$\bra q_\perp(\bmath{k},w_1) q^*_\perp(\bmath{k},w_2)\ket_\mathrm{4pt}$, which is of order $\upsilon^2/c^2$ and thus 
strongly suppressed. The derivation of the term is easy prey: It can be done in complete analogy to that of the Ostriker-Vishniac 
effect \citep{1986ApJ...306L..51O, 1987ApJ...322..597V}, where any optical depth depending on redshift needs to be replaced by 
the appropriate weighting function (c.f. Sect.~\ref{sect_ctau}) and conversions from dark matter densities into baryonic densites 
are to be discarded.

The derivation evolves cross-terms between the velocity and density fields, perhaps the most elegant way of reducing it to a sum 
of 2-point correlations is given by \citet{2002PhRvL..88u1301M}, using a result from \citet{monin_yaglom_1,monin_yaglom_2}:
\begin{displaymath}
(2\pi)^3\bra q_i(\bmath{k}) q^*_j(\bmath{k})\ket_\mathrm{4pt} \equiv P_{qq}^{ij}(\left|\bmath{k}\right|)\simeq
\end{displaymath}
\begin{displaymath}
\quad \frac{4}{c^2}\int\frac{\dd^3 p}{(2\pi)^3} \int\frac{\dd^3 p^\prime}{(2\pi)^3}
(2\pi)^3\delta_D(\bmath{k}-\bmath{p}-\bmath{p}^\prime)
\end{displaymath}
\begin{equation}
\quad\times
\left[
\frac{\bmath{p}^i}{\left|\bmath{p}^i\right|}\frac{\bmath{p}^j}{\left|\bmath{p}^j\right|} 
P_{\upsilon\upsilon}(\left|\bmath{p}\right|) P_{\delta\delta}(\left|\bmath{p}^\prime\right|) +
\frac{\bmath{p}^i}{\left|\bmath{p}^i\right|}\frac{\bmath{p}^{\prime j}}{\left|\bmath{p}^{\prime j}\right|}
P_{\delta\upsilon}(\left|\bmath{p}\right|) P_{\delta\upsilon}(\left|\bmath{p}^\prime\right|) 
\right]\mbox{,}
\label{eqn_pij}
\end{equation}
where the irreducible 4-point correlation $P_{\delta\upsilon\delta\upsilon}(\bmath{k})$ has been neglected. 

Following \citet{2002PhRvL..88u1301M}, the projection to be carried out is $(2\pi)^3 \bra q_\perp(\bmath{k}) 
q^*_\perp(\bmath{k})\ket_\mathrm{4pt} = 2\sum_{ij} \bmath{e}_i\bmath{e}_j P_{qq}^{ij}(\left|\bmath{k}\right|)$, where 
$\bmath{e}_i$ and $\bmath{e}_j$ are unit vectors along the lines-of-sight. The expression for 
$P_{qq}^{ij}(\left|\bmath{k}\right|)$ is given by eqn.~(\ref{eqn_pij}). In neglecting the irreducible 4-point term one 
obtains:
\begin{equation}
(2\pi)^3 \bra q_\perp(\bmath{k}) q^*_\perp(\bmath{k})\ket_\mathrm{4pt} = 
\frac{1}{c^2}\int\frac{\dd^3 p}{(2\pi)^3}\left\{\vphantom{\int}\right.
\end{equation}
\begin{displaymath}
\quad(1-x^2) 
P_{\delta\delta}(\left|\bmath{k}-\bmath{p}\right|) P_{\upsilon\upsilon}(\left|\bmath{p}\right|)
- \frac{(1-x^2) p}{\left|\bmath{k}-\bmath{p}\right|} 
P_{\delta\upsilon}(\left|\bmath{k}-\bmath{p}\right|) P_{\delta\upsilon}(\left|\bmath{p}\right|)
\left.\vphantom{\int}\right\}
\end{displaymath}
Inserting the time-evolution of the density-velocity and velocity-velocity cross correlation terms,
\begin{eqnarray}
\bra\upsilon_\perp(\bmath{k},w_1)\upsilon^*_\perp(\bmath{k},w_2)\ket & = &
g^\prime(w_1) g^\prime(w_2) \bra\upsilon_\perp(\bmath{k})\upsilon^*_\perp(\bmath{k})\ket
\mbox{,} \\
\bra\upsilon_\perp(\bmath{k},w_1)\delta^*(\bmath{k},w_2)\ket & = &
g^\prime(w_1) D(w_2) \bra\upsilon_\perp(\bmath{k})\delta^*(\bmath{k})\ket
\mbox{,}
\end{eqnarray}
yields the final result:
\begin{displaymath}
(2\pi)^3 \bra q_\perp(\bmath{k},w_1) q^*_\perp(\bmath{k},w_2)\ket_\mathrm{4pt} = 
4 D(w_1) D(w_2) g^\prime(w_1) g^\prime(w_2) \times
\end{displaymath}
\begin{equation}
\quad\frac{2\pi}{c^2} \int\dd p\:\int_{-1}^{+1}\clm\dd x P(\left|\bmath{k-p}\right|) P(\left|\bmath{p}\right|)
\frac{k(1-x^2)(k-2xp)}{k^2-2xkp+p^2}\mbox{.}
\end{equation}

% ============================================================================================== %
\subsection{Corrections to the power spectrum}
\begin{figure}
\resizebox{\hsize}{!}{\includegraphics{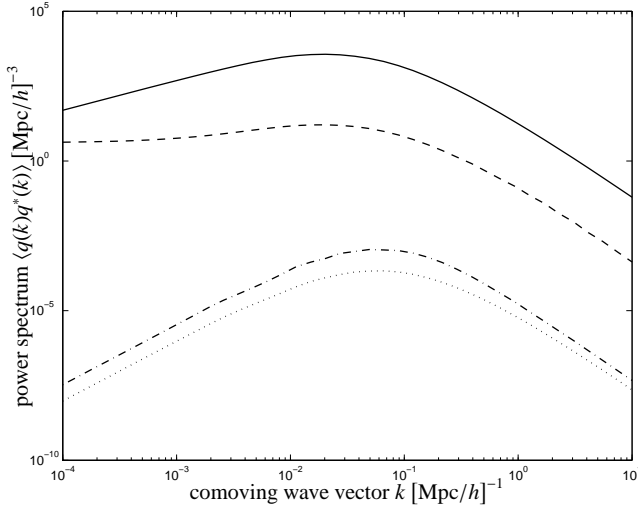}}
\caption{Three-dimensional power spectrum $\bra q_\perp(\bmath{k})q^*_\perp(\bmath{k})\ket$ including dark matter 
currents  perpendicular to the line-of-sight, split up into the 2-point contribution (solid line), the 3-point contribution 
(dashed line) and the 4-point contribution (dash-dotted line). Additionally, the 4-point term of the currents 
parallel to the line-of-sight $\bra q_\parallel(\bmath{k}) q^*_\parallel(\bmath{k})\ket$ is drawn (dotted line). The power 
spectra are given for the present epoch, i.e. $a=1$ and $z=0$.}
\label{fig_gravitomagnetic_3d}
\end{figure}

The three-dimensional power spectra $\bra q_\perp(\bmath{k}) q^*_\perp(\bmath{k})\ket$ of the matter currents parallel to the 
line-of-sight is shown in Fig.~\ref{fig_gravitomagnetic_3d} for the various $n$-point contributions. Compared to the dominating 
2-point term, the 3-point term is smaller by more than two orders of magnitude on small scales, but it becomes important on large 
spatial scales beyond 1~Gpc where it levels out. On these large scales, however, limitations due to cosmic variance and 
insufficient sampling due to galactic foregrounds cast doubt on its detectability. The leveling on large spatial scales of the 
3-point term is due to the fact, that for small $k$ all powers in $p$ in eqn.~(\ref{eqn_threept}) add up to zero, which results in 
a flat curve for $\bra q_\perp(\bmath{k}) q^*_\perp(\bmath{k})\ket_\mathrm{3pt}$. In comparison to the 3-point term, the 4-point 
term is smaller by another three orders of magnitude, but in shape it very much resembles the 2-point term and its influence on 
the weak lensing power spectrum is safely negligible. 

% ============================================================================================== %
\subsection{Projected lensing power spectra}
The final expression for $\bra q_\perp(\bmath{k}) q_\perp^*(\bmath{k})\ket$ can be projected by means of 
eqn.~(\ref{eqn_proj_perp}) to yield the angular power spectrum of any lensing quantity, for example the convergence 
$\kappa$. The distance weighting function to be employed can be read off from eqn.~(\ref{eqn_kappa_cosmo}):
\begin{equation}
W_\mathrm{L}(w) = \frac{3 H_0^2\Omega_0}{2c^2}\frac{f_K(w)}{a(w)}
\int_w^{w_\mathrm{max}}\cll\dd w^\prime Z(w^\prime)\frac{f_K(w^\prime-w)}{f_K(w^\prime)}\mbox{.}
\end{equation}
By substituting $y=kw$, the distance weighting $W_\mathrm{L}(w)$ can be combined with the time evolution of the correlators to 
yield the functions
\begin{equation}
\varphi_\ell(k)_\mathrm{2pt} =
\left[\int_0^{y_\mathrm{max}}\cll\dd y W_\mathrm{L}\left(\frac{y}{k}\right)\frac{J_l(y)}{y} D(y)\right]^2
\mbox{,}
\end{equation}
\begin{equation}
\varphi_\ell(k)_\mathrm{3pt} = 
\int_0^{y_\mathrm{max}}\cll\dd y W_\mathrm{L}\left(\frac{y}{k}\right)\frac{J_l(y)}{y} D(y) G(y)
\int_0^{y_\mathrm{max}}\cll\dd y W_\mathrm{L}\left(\frac{y}{k}\right)\frac{J_l(y)}{y} D(y)
\end{equation}
\begin{equation}
\varphi_\ell(k)_\mathrm{4pt} =
\left[\int_0^{y_\mathrm{max}}\cll\dd y W_\mathrm{L}\left(\frac{y}{k}\right)\frac{J_l(y)}{y} D(y) G(y)\right]^2
\mbox{,}
\end{equation}
which carry out the projection of the 3-dimensional power spectrum $\bra q_\perp(k) q^*_\perp(k)\ket$ to the angular power 
spectrum $C_\kappa(\ell)$ by convolution:
\begin{equation}
C_\kappa(\ell) = \frac{1}{(2\pi)^2}\ell(\ell+1)\int\dd k\: \bra q_\perp(k) q^*_\perp(k)\ket\times\varphi_\ell(k) \mbox{,}
\end{equation}
where the associativity of the time-evolution enables the 3-fold integration in eqn.~(\ref{eqn_proj_perp}) to be reduced to a 
2-fold integration. 

\begin{figure}
\resizebox{\hsize}{!}{\includegraphics{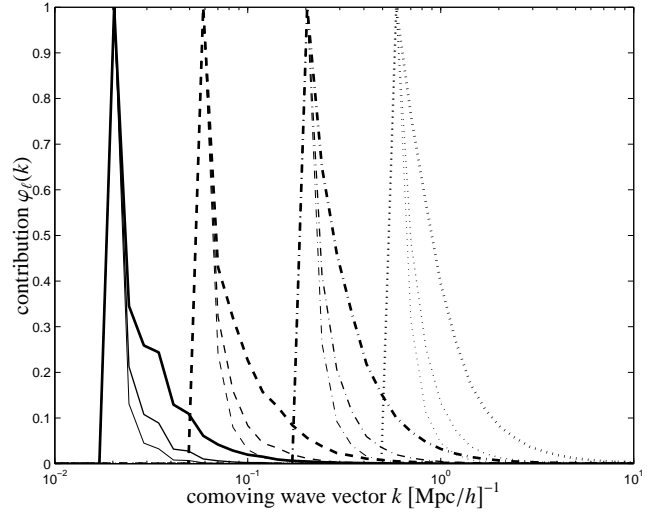}}
\caption{Contribution $\varphi_\ell(k)$ of the 2-point terms (thick lines), 3-point terms (medium lines) and 4-point terms 
(thin lines) to the angular power spectrum $C_\kappa(\ell)$ of the weak lensing convergence $\kappa$ as a function of wave vector 
$k$, for $\ell=100$ (solid line), $\ell=300$ (dashed line), $\ell=1000$ (dash-dotted line) and $\ell = 3000$ (dotted line). The 
curves are normalised to unity.}
\label{fig_gravitomagnetic_contrib}
\end{figure}

The functions $\varphi_\ell(k)_\mathrm{2pt}$, $\varphi_\ell(k)_\mathrm{3pt}$ and $\varphi_\ell(k)_\mathrm{4pt}$ are shown in 
Fig.~\ref{fig_gravitomagnetic_contrib}. Clearly, the fluctuations on a certain angular scale described by the angular power 
spectrum $C(\ell)$ are dominated by spatial fluctuations with a certain wave vector $k$, which leads the peak of the function 
$\varphi_\ell(k)$ to shift with increasing $\ell$. The projection kernels $\varphi_\ell(k)$ for the different $n$-point 
correlation functions show the common feature of rising fast at small $k$, but their decays at large $k$ vary appreciably, because 
the increasing influence of the time evolution of the velocities $G(w)$ makes the functions to drop faster with increasing values 
of $k$. In this way, the observed asymptotic behaviour is $\varphi_\ell(k_\mathrm{2pt})\propto k^{-2}$ for the 2-point projector, 
but the $\varphi_\ell(k_\mathrm{3pt})$ and $\varphi_\ell(k_\mathrm{4pt})$ exhibit faster decays that are not described by a 
mere power law. 

\begin{figure}
\resizebox{\hsize}{!}{\includegraphics{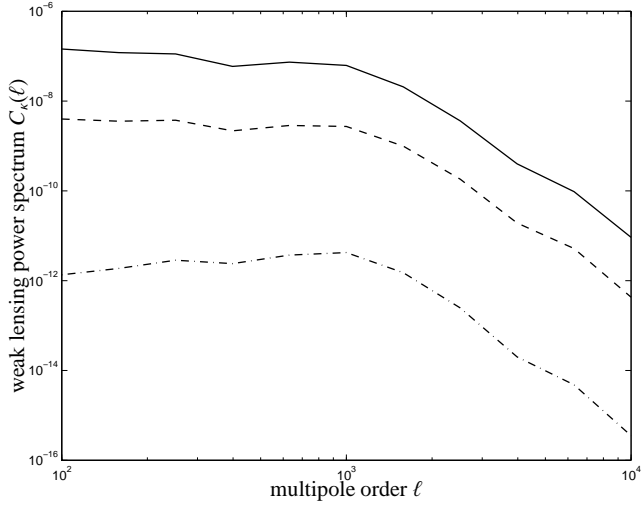}}
\caption{Angular power spectrum $C_\kappa(\ell)$ of the weak lensing convergence $\kappa$ and its correction due to 
gravitomagnetic potentials. The contributions from the 2-point term (solid line), the 3-point term (dashed line) and the 
4-point term (dash-dotted line) are given separately.}
\label{fig_gravitomagnetic_2d}
\end{figure}

The angular power spectrum of the weak lensing convergence $C_\kappa(\ell)$ and its corrections due to gravitomagnetic terms is 
depicted in Fig.~\ref{fig_gravitomagnetic_2d}. Even at the largest angular scales considered here, the function $\varphi_\ell(k)$ 
peaks at values of $k$ at which the corrections of the 3-point term are negligible. The detection of corrections to the weak 
lensing power spectrum due to gravitomagnetic terms requires the measurement of weak lensing shear on very large angular scales, 
which is beyond feasibility with current technology. On large angular scales, cosmic variance additionally limits the 
observability of gravitomagnetic lensing.

% ############################################################################################## %
% ------------------------------ section: rees sciama effect ----------------------------------- %
\section{Integrated Sachs-Wolfe effect}\label{sect_rees_sciama}

% ============================================================================================== %
\subsection{Definitions}
The growth of structure imprints additional anisotropies on the cosmic microwave background (CMB) by the time variation of 
the gravitational potentials along the propagation path of a CMB photon. This effect is called the integrated Sachs-Wolfe (iSW)
effect in the regime of linear structure formation \citep{1967ApJ...147...73S, 1994PhRvD..50..627H} and Rees-Sciama effect 
\citep{rees_sciama_orig, 1996ApJ...460..549S, 2002PhRvD..65h3518C} if the density perturbations grow nonlinearly. The 
perturbations $\Delta T$ of the sky temperature $T$ can be written as:
\begin{equation}
\tau\equiv\frac{\Delta T}{T} = -\frac{2}{c^3} \int \dd w\: \frac{\partial\Phi}{\partial t}\mbox{,}
\end{equation}
where $\partial\Phi/\partial t$ is the derivative of the gravitational potentials with respect to conformal time $t$. Similar to 
gravitomagnetic lensing discussed in Sect.~\ref{def_gravitomagnetic} (c.f. eqns.~(\ref{eqn_def_neff}) and 
(\ref{eqn_lensing_alpha})), the effect is of the order $1/c^3$.

% ============================================================================================== %
\subsection{Connection to the gravitomagnetic potentials}
Using the definition of $\Phi(\bmath{r})$ and the equation of continuity, $\dot{\rho} + \mathrm{div}\bmath{j}$, where 
$\bmath{j} = \rho\bmath{\upsilon}$ is the matter current density, one obtains for the time derivative of $\Phi$:
\begin{equation}
\frac{\partial}{\partial t}\Phi(\bmath{r},t) =
-G\int\dd^3r^\prime\frac{\dot{\rho}(\bmath{r}^\prime)}{\left|\bmath{r}-\bmath{r}^\prime\right|} = 
G\int\dd^3r^\prime\frac{\nabla^\prime\cdot\bmath{j}(\bmath{r}^\prime)}{\left|\bmath{r}-\bmath{r}^\prime\right|}\mbox{.}
\end{equation}
Integration by parts with respect to $\dd^3r^\prime$ yields:
\begin{equation}
G\int\dd^3r^\prime\frac{\nabla^\prime\cdot\bmath{j}(\bmath{r}^\prime)}{\left|\bmath{r}-\bmath{r}^\prime\right|} = 
-G\int\dd^3r^\prime \bmath{j}(\bmath{r}^\prime)\cdot\nabla^\prime\frac{1}{\left|\bmath{r}-\bmath{r}^\prime\right|}\mbox{.}
\end{equation}
With the identity
\begin{equation}
\nabla^\prime\frac{1}{\left|\bmath{r}-\bmath{r}^\prime\right|} = -\nabla\frac{1}{\left|\bmath{r}-\bmath{r}^\prime\right|},
\end{equation}
the derivative with respect to the primed coordinate can be replaced by a derivative with respect to the unprimed one. In the 
last steps, the $\nabla$-operator can be drawn in front of the integral and the definition of $\bmath{A}$ (c.f. 
eqn.~(\ref{eqn_lensing_defa})) be inserted to yield:
\begin{equation}
G\int\dd^3r^\prime\bmath{j}(\bmath{r}^\prime)\cdot\nabla\frac{1}{\left|\bmath{r}-\bmath{r}^\prime\right|} = 
-\nabla\cdot\left(-G\int\dd^3r^\prime\frac{\bmath{j}(\bmath{r}^\prime)}{\left|\bmath{r}-\bmath{r}^\prime\right|}\right) = 
-\mathrm{div}\bmath{A}\mbox{.}
\end{equation}
Thus, the final result reads:
\begin{equation}
\frac{\partial}{\partial t}\Phi(\bmath{r},t) = -\mathrm{div}\bmath{A}\rightarrow
\tau = \frac{2}{c^3}\int\dd w\: \mathrm{div}\bmath{A}.
\label{eqn_rees_sciama_diva}
\end{equation}
Eqn.~(\ref{eqn_rees_sciama_diva}) has a lucid interpretation: The formation of objects such as clusters of galaxies requires 
the matter fluxes $\bmath{j}$ to converge and to accumulate matter ($\dot\rho>0$). Consequently, potential wells deepen and 
give rise to the iSW-effect in regions where $\mathrm{div}\bmath{A}$ does not vanish. The iSW-effect measures the rate of change 
of a potential with respect to conformal time, or equivalently, the divergence of the vector potential $\bmath{A}$, which is 
proportional to the accretion rate in the Lagrangian frame.

% ============================================================================================== %
\subsection{Putting the Sachs-Wolfe effect in a cosmological context}
In order to relate the statistical properties of the iSW temperature fluctuations $\tau(\bmath{\theta})$ to 
those of the matter streams $\bmath{j}(\bmath{r})$, the auxiliary vector field $\bmath{\chi}(\bmath{\theta})$ is introduced, which 
is  defined as the negative gradient of the iSW temperature fluctuation $\tau(\bmath{\theta})$:
\begin{equation}
\bmath{\chi}(\bmath{\theta}) \equiv -\nabla\tau(\bmath{\theta})\mbox{,}
\label{eqn_chi_def}
\end{equation}
i.e. $\bmath{\chi}(\bmath{\theta})$ points along the steepest descent in temperature from hot to cold patches in an iSW field. 
Inserting eqn.~(\ref{eqn_rees_sciama_diva}) into the defining equation for $\bmath{\chi}(\bmath{\theta})$ and converting the 
derivation with respect to the angular variable $\bmath{\theta}$ into a derivation with respect to the comoving variable 
$\bmath{r}$ by using $\nabla_\theta= f_K(w)\nabla$, enables interchanging integration and differentiation:\begin{equation}
\bmath{\chi}(\bmath{\theta}) 
= \frac{2}{c^3}\int\dd w\: f_K(w)\:\nabla~(\mathrm{div}\bmath{A}) 
= \frac{2}{c^3}\int\dd w\: f_K(w)\:\Delta\bmath{A}\mbox{.}
\end{equation}
Additionally, the replacement $\nabla~(\mathrm{div}\bmath{A}) = \Delta\bmath{A}$ is inserted, which is valid if 
$\mathrm{rot}~\mathrm{rot}\bmath{A}=\bmath{0}$. This is fulfilled in vorticity-free velocity fields, 
$\mathrm{rot}\bmath{j}=\bmath{0}$. In linear theory, initial vorticity perturbations are damped and after a sufficiently long 
time, the linear velocity field should be completely irrotational. Even in the regime of quasi- or nonlinear structure 
formation, Kelvin's circulation theorem forces the flow to remain irrotational and described by a velocity potential
until dissipative processes on smallest scales give rise to vortical flows.

Inserting Laplace's equation in the comoving frame, $\Delta\bmath{A} = 4\pi G a^2 \bra\rho\ket(\delta\bmath{\upsilon})$ with 
the source term $\bmath{j}=\delta\bmath{\upsilon}$, allows to replace Newton's constant $G$ and the ambient mass 
density $\bra\rho\ket$ by using $\rho_\mathrm{crit} = 3 H_0^2/(8\pi G)$, $\bra\rho\ket_0=\Omega_0\rho_\mathrm{crit}$ and 
$\bra\rho\ket=\bra\rho\ket_0 / a^3$:
\begin{equation}
\bmath{\chi}(\bmath{\theta}) 
= \frac{2}{c^3}\int\dd w\: f_K(w)\:\frac{4\pi G \bra\rho\ket}{a}\bmath{j} 
= \frac{3 H_0^2 \Omega_0}{c^2}\int\dd w \frac{f_K(w)}{a(w)}\: \frac{\bmath{j}}{c}\mbox{.}
\label{eqn_rs_cosmo}
\end{equation}
The structural similarity of eqn.~(\ref{eqn_rs_cosmo}) with the weak lensing convergence eqn.~(\ref{eqn_kappa_cosmo}) is 
striking.

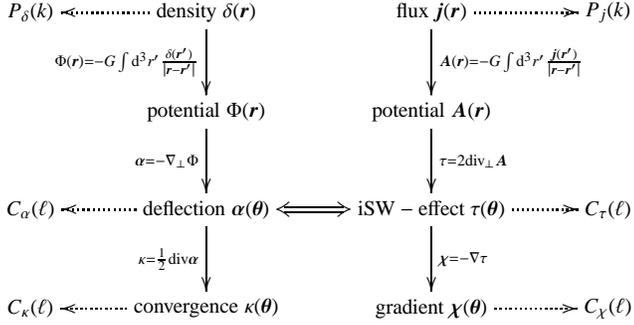
\begin{figure}
\begin{center}
\begin{displaymath}
\xymatrix{
P_\delta(k)
& \ar@{.>}[l] \mathrm{density}~\delta(\bmath{r})
\ar[d]_{\Phi(\bmath{r})=-G\int\dd^3 r^\prime\frac{\delta(\bmath{r}^\prime)}{\left|\bmath{r}-\bmath{r}^\prime\right|}}
& \ar[d]^{\bmath{A}(\bmath{r})=-G\int\dd^3 r^\prime\frac{\bmath{j}(\bmath{r}^\prime)}{\left|\bmath{r}-\bmath{r}^\prime\right|}}
\mathrm{flux}~\bmath{j}(\bmath{r}) \ar@{.>}[r] & P_j(k) \\
& \mathrm{potential}~\Phi(\bmath{r}) \ar[d]_{\bmath{\alpha}=-\nabla_\perp\Phi} & 
\ar[d]^{\tau=2\mathrm{div}_\perp\bmath{A}} \mathrm{potential}~\bmath{A}(\bmath{r}) & \\
C_\alpha(\ell) & \ar@{.>}[l] 
\mathrm{deflection}~\bmath{\alpha}(\bmath{\theta}) \ar[d]_{\kappa=\frac{1}{2}\mathrm{div}\bmath{\alpha}}
\ar@{<=>}[r]
& \ar[d]^{\bmath{\chi}=-\nabla\tau} \mathrm{iSW-effect}~\tau(\bmath{\theta}) \ar@{.>}[r] & C_\tau(\ell) \\
C_\kappa(\ell) & \ar@{.>}[l] \mathrm{convergence}~\kappa(\bmath{\theta})
& \mathrm{gradient}~\bmath{\chi}(\bmath{\theta}) \ar@{.>}[r] & C_\chi(\ell)
}
\end{displaymath}
\end{center}
\caption{Flow chart with correspondent quantities in gravitational lensing (left column) and the integrated Sachs-Wolfe effect 
(right column). Thequantity analogous to the iSW temperature fluctuation $\tau(\bmath{\theta})$ in weak gravitational lensing is 
the deflection angle $\bmath{\alpha}(\bmath{\theta})$.}
\label{fig_comparison}
\end{figure}

Now, the 2-point correlation of the iSW temperature gradient field $\bmath{\chi}(\bmath{\theta})$ is related 
to the matter flux density $\bmath{j}(\bmath{r})$. For the derivation of the correlation function $C_\tau(\ell)$ of the 
temperature field $\tau(\bmath{\theta})$ itself, one rewrites eqn.~(\ref{eqn_chi_def}) in Fourier space, yielding:
\begin{equation}
\bmath{\chi}(\bmath{\theta}) = \int\frac{\dd^2\ell}{(2\pi)^2}\:\bmath{\chi}(\bmath{\ell}) \exp(i\bmath{\ell}\cdot\bmath{\theta})
\rightarrow\bmath{\chi}(\bmath{\ell}) = i \bmath{\ell} \tau(\ell)
\end{equation}
The expansion into Fourier modes rather than spherical harmonics is permissible, because $\tau$ is expected to show 
fluctuations on small angular scales, so that $\tau$ can be considered on a plane locally tangential to the celestial sphere. 
Squaring immediately gives:
\begin{equation}
C_\chi(\ell) = \ell^2 C_\tau(\ell)\simeq \ell(\ell+1) C_\tau(\ell)\mbox{,}
\end{equation}
where the last step is a valid approximation for small angular scales. The complementarity of gravitational lensing and the 
iSW-effect and the lensing counterparts of iSW quantities are illustrated in the flow chart Fig.~\ref{fig_comparison}.

% ============================================================================================== %
\subsection{Perturbative treatment}
In the following we adopt the approximation that the rate of change of a potential is constant during the photon passage and 
that the accretion geometry does not change significantly. In this approximation, the component $\dd A_z/\dd z$ of 
$\mathrm{div}\bmath{A}$ is cancelled by the integration in eqn.~(\ref{eqn_rees_sciama_diva}) and makes the iSW 
effect to measure the components of $\mathrm{div}\bmath{A}$ perpendicular to the line-of-sight, i.e. 
$\tau\propto\mathrm{div}_\perp\bmath{A}=\dd A_x/\dd x + \dd A_y/\dd y$. Consequently, the matter fluxes 
$\bmath{q}_\perp(\bmath{x}) = \bmath{j}_\perp(\bmath{x})/c = \delta(\bmath{x}) \bmath{\upsilon}_\perp(\bmath{x})/c$ 
perpendicular to the line-of-sight primarily give rise to the iSW-effect. Accordingly, the fluctuations in the CMB due to the 
Rees-Sciama effect, being sensitive to the components of the matter flux perpendicular to the line-of-sight, are dominated by the 
components of $k$-modes parallel to the line of sight.

Power spectra of the form $\bra q_\parallel(\bmath{k}) q^*_\parallel(\bmath{k})\ket$  have been considered by many authors in the 
derivation of the Ostriker-Vishniac effect \citep[e.g.][]{1987ApJ...322..597V, 1998PhRvD..58d3001J}. In order to obtain the 
projection onto the line-of-sight, $(2\pi)^3\bra q_\parallel(\bmath{k}) q^*_\parallel(\bmath{k})\ket_\mathrm{4pt} = \sum_{ij}
\frac{\bmath{k}^i}{\left|\bmath{k}^i\right|}\frac{\bmath{k}^j}{\left|\bmath{k}^j\right|} P_{qq}^{ij}(\left|\bmath{k}\right|)$, 
has to be carried out, which can be interpreted as the quadratic form $\hat{\bmath{k}}\vphantom{k}^T\tilde{P}\hat{\bmath{k}}$ 
with a unit vector $\hat{\bmath{k}}$ and the matrix $\tilde{P} = P^{ij}_{qq}$ (compare eqn.~\ref{eqn_pij}). The matrix $\tilde{P}$ 
introducing the scalar product $\hat{\bmath{k}}\vphantom{k}^T\tilde{P}\hat{\bmath{k}}$ is positive definite, due to the reality of 
the density and velocity fields.
\begin{equation}
(2\pi)^3\bra q_\parallel(\bmath{k}) q^*_\parallel(\bmath{k})\ket_\mathrm{4pt} = 
\end{equation}
\begin{displaymath} 
\frac{4}{c^2}\int\frac{\dd^3 p}{(2\pi)^3} 
x^2 
P_{\delta\delta}(\left|\bmath{k}-\bmath{p}\right|) P_{\upsilon\upsilon}(\left|\bmath{p}\right|) +
\frac{(k - px) x}{\left|\bmath{k}-\bmath{p}\right|} 
P_{\delta\upsilon}(\left|\bmath{k}-\bmath{p}\right|) P_{\delta\upsilon}(\left|\bmath{p}\right|)
\end{displaymath}
The scalar product $\bmath{p}\bmath{k}$ is again equal to $p k x$, where $x$ is the cosine of the angle of separation. 
Inserting the velocity-density and velocity-velocity cross correlation functions with their proper time evolution,
\begin{eqnarray}
\bra\upsilon_\parallel(\bmath{k},w_1)\delta^*(\bmath{k},w_2)\ket & = &
g^\prime(w_1) D(w_2) \bra\upsilon_\parallel(\bmath{k})\delta^*(\bmath{k})\ket
\mbox{,} \\
\bra\upsilon_\parallel(\bmath{k},w_1)\upsilon_\parallel^*(\bmath{k},w_2)\ket & = &
g^\prime(w_1) g^\prime(w_2) \bra\upsilon_\parallel(\bmath{k})\upsilon_\parallel^*(\bmath{k})\ket
\mbox{,}
\end{eqnarray}
yields the final result:
\begin{displaymath}
(2\pi)^3\bra q_\parallel(\bmath{k},w_1) q_\parallel(\bmath{k},w_2)\ket_\mathrm{4pt} =
D(w_1) D(w_2) g^\prime(w_1) g^\prime(w_2) \times
\end{displaymath}
\begin{equation}
\quad \frac{2\pi}{c^2}\int\dd p \int_{-1}^{+1}\clm\dd x P(\left|\bmath{k-p}\right|) P(\left|\bmath{p}\right|)
\frac{kx(kx-2px^2+p)}{k^2-2xkp+p^2}\mbox{.}
\end{equation}

% ============================================================================================== %
\subsection{Power spectrum of dark matter currents}
The three-dimensional power spectrum $\bra q_\parallel(\bmath{k}) q^*_\parallel(\bmath{k})\ket$ of the matter currents 
perpendicular to the line-of-sight is given in Fig.~\ref{fig_gravitomagnetic_3d}. Its amplitude is by a factor of 4 smaller than 
the power spectrum $\bra q_\perp(\bmath{k}) q^*_\perp(\bmath{k})\ket$, because the iSW-effect measures the streams 
$\delta\upsilon$ in contrast to gravitomagnetic lensing, where the source terms read $(1+2\upsilon/c)\delta$. Despite the fact 
that different projections are considered, the shape and asymptotic forms of $\bra q_\parallel(\bmath{k}) 
q^*_\parallel(\bmath{k})\ket$ and $\bra q_\perp(\bmath{k}) q^*_\perp(\bmath{k})\ket$ are very similar.

% ============================================================================================== %
\subsection{integrated Sachs-Wolfe angular power spectrum}\label{sect_ctau}
\begin{figure}
\resizebox{\hsize}{!}{\includegraphics{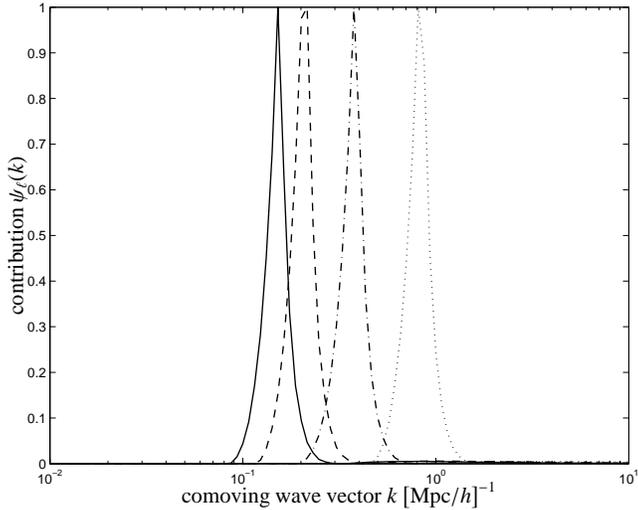}}
\caption{Contribution $\psi_\ell(k)$ of the 4-point term to the angular power spectrum $C_\tau(\ell)$ of the iSW temperature 
fluctuations $\tau$ as a function of wave vector $k$, for $\ell=100$ (solid line), $\ell=300$ (dashed line), $\ell=1000$ 
(dash-dotted line) and $\ell = 3000$ (dotted line). The curves have been normalised to a peak value of unity.}
\label{fig_rees_sciama_contrib}
\end{figure}

\begin{figure}
\resizebox{\hsize}{!}{\includegraphics{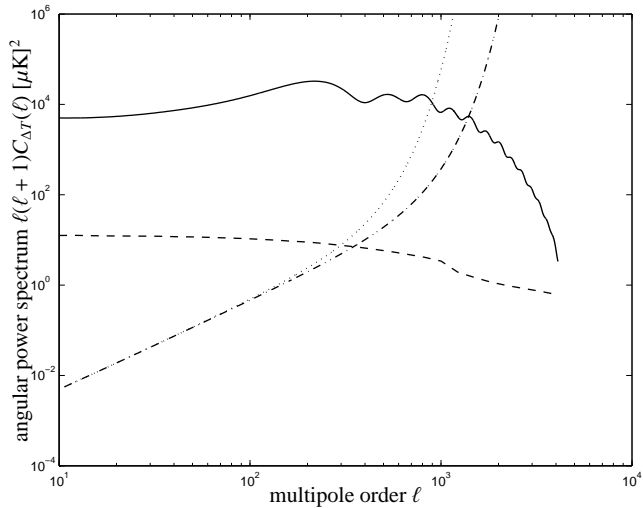}}
\caption{Angular power spectrum $C_{\Delta T}(\ell) = T_\mathrm{CMB}^2 C_\tau(\ell)$ of the iSW temperature fluctuations 
$\tau(\bmath{\theta})$ (dashed line). The CMB power spectrum $C_\mathrm{CMB}(\ell)$ for the $\Lambda$CDM model (solid line) 
and the limiting {\em Planck}-sensitivies $C_\mathrm{noise}(\ell)$ for angular resolutions $\Delta\theta=5\farcm0$ (dash-dotted 
line) and $\Delta\theta=9\farcm1$ (dotted line) are depicted for comparison.}
\label{fig_rees_sciama_2d}
\end{figure}

Fig.~\ref{fig_rees_sciama_2d} shows the angular power spectra $C_\tau(\ell)$ of the iSW-effect 
$\tau(\bmath{\theta})$ and $C_\chi(\ell)$ of the iSW temperature gradient $\bmath{\chi}(\bmath{\theta})$ which 
have been obtained by applying the projection formula~(\ref{eqn_proj_parallel}) to the spectrum $\bra q_\parallel(\bmath{k}) 
q_\parallel^*(\bmath{k})\ket$ with the weighing function
\begin{equation}
W_\mathrm{iSW}(w) = \frac{3 H_0^2 \Omega_0}{c^2} \frac{f_K(w)}{a(w)}\mbox{,}
\end{equation}
which can be read off from eqn.~(\ref{eqn_rs_cosmo}). The redshift-weightings and the time-evolution of the density and velocity 
fields can be combined, which yields the function (\ref{eqn_rs_psi}) after substituting $y=kw$,
\begin{equation}
\psi_\ell(k)_\mathrm{4pt} =
\left[\int_0^{y_\mathrm{max}}\cll\dd y W_\mathrm{iSW}\left(\frac{y}{k}\right)\frac{\dd J_l(y)}{\dd y} D(y) G(y)\right]^2
\label{eqn_rs_psi}
\end{equation}
which mediates between the 3-dimensional power spectrum $\bra q_\parallel(k) q^*_\parallel(k)\ket$ and the angular power spectrum 
$C_\tau(\ell)$ by convolution:
\begin{equation}
C_\tau(\ell) = \frac{1}{(2\pi)^2}\int\dd k\: \bra q_\parallel(k) q^*_\parallel(k)\ket\times\psi_\ell(k)\mbox{.}
\end{equation}
Again, the 3-fold integration in eqn.~(\ref{eqn_proj_parallel}) is reduced to a 2-fold integration. The shape of the function 
$\psi_\ell(k)$ is depicted in Fig.~\ref{fig_rees_sciama_contrib} for various values of $\ell$. In contrast to the function 
$\varphi_\ell(k)$ used in the projection of the lensing power spectra, the function $\psi_\ell(k)$ is symmetric about its peak, 
which is caused by the replacement of $J_\ell(y)/y$ with the derivative $\dd J_\ell(y)/\dd y$. The fast variability is again due 
to the strong influence of the velocity time evolution $G(y)$.

The angular power spectrum $C_\tau(\ell)$ of the iSW temperature fluctuations $\tau(\bmath{\theta})$ along with the primary CMB 
fluctuations and the limiting {\em Planck}-sensitivity is depicted in Fig.~\ref{fig_rees_sciama_2d}. The angular power spectrum 
has an amplitude of $\simeq 3\times 10^{-11}~\mathrm{K}^2$ at small $\ell$ and shows a slow variation with the multipole order 
$\ell$. The amplitude agrees well with the result from \citet{1996ApJ...460..549S}, but the decline of the power spectrum on large 
angular scales could not be confirmed, which is due to the fact that for large angles, the Bessel functions $J_l(x)$ are a poor 
approximation to the Legendre polynomials $P_\ell(x)$. The position of the peak in the projection kernel $\psi_\ell(k)$ suggests 
that on the largest scales considered here, the angular spectrum $C_\tau(\ell)$ is dominated by fluctuations at the maximum of 
$P(k)$ on scales at $k^{-1}\simeq 10$~Mpc. With increasing multipole order $\ell$, the peak in $\psi_\ell(k)$ shifts only slowly 
towards higher values of $k$, which explains the small variation of $C_\chi(\ell)=\ell(\ell+1) C_\tau(\ell)$.

The channel averaged {\em Planck}-sensitivy is described by \citep{1995PhRvD..52.4307K, 1996MNRAS.281.1297T}:
\begin{equation}
C_\mathrm{noise}(\ell) = \frac{4\pi\sigma^2}{N_\mathrm{pix}}\exp\left[\theta_b^2 \ell(\ell+1)\right]\mbox{,}
\end{equation}
where $N_\mathrm{pix}\simeq 5.03\times10^7$ is the number of pixels and $\theta_b$ the FWHM extension of the {\em Planck}-beam. 
For the average amplitude of the noise $\sigma_\mathrm{eff}$ per solid angle subtended by a single pixel we use the quadratic 
harmonic mean over all {\em HFI}-channels:
\begin{equation}
\frac{1}{\sigma_\mathrm{eff}^2} = \sum_{i=1}^{6}\frac{1}{\sigma_i^2}
\longrightarrow\sigma_\mathrm{eff} = 13.42~\mu\mathrm{K}\mbox{.}
\end{equation}

The sensitiviy considerations suggest that the iSW-effect is well above the noise level of the combined {\em Planck HFI}-channels, 
so that the power spectrum of $C_\tau(\ell)$ should be observable for angular scales $\ell\lsim200$ as a contribution to the 
primary CMB fluctuations $C_\mathrm{CMB}(\ell)$, which in Fig.~\ref{fig_rees_sciama_2d} have been computed using the {\tt CMBfast} 
code by \citet{1996ApJ...469..437S}. The signal can of course be amplified by cross-correlation with a suitable population of tracer galaxies.

% ############################################################################################## %
% ------------------------------ section: Summary ---------------------------------------------- %
\section{Summary}\label{sect_summary}
The scope of this paper is to derive the corrections to the power spectrum of weak gravitational lensing due to 
gravitomagnetic terms in the metric by perturbation theory. Within the same formalism, the power spectrum of the iSW-effect can be 
determined as well.

\begin{itemize}
\item{The iSW-effect and gravitomagnetic lensing measure the evolution of velocities and densities in the 
large-scale structure and are sensitive to the cosmological parameters $\Omega_\mathrm{M}$ and $\sigma_8$. Applied to single 
objects like clusters, where the above described formalism equally applies, the iSW-effect would allow to 
measure the cosmological evolution of merger rates and dark matter accretion strengths \citep{2002MNRAS.331...98V, 
2002ApJ...568...52W, 2003MNRAS.339...12Z}.}

\item{Gravitomagnetic lensing would test general relativity on the largest scales (Mpc - Gpc) to second order in $\upsilon/c$, and could help to decide in favour of or against other metric theories of gravity. It should be emphasised that in the current theoretical description of structure formation or in current numerical simulations the motion of bodies is described by classical 
mechanics, i.e. instantaneous propagation of potentials and no relativistic increase of inertial mass with velocity, but the 
interaction of light with matter should be treated in the framework of the post-Newtonian limits of general relativity.}

\item{Gravitomagnetic terms influence the weak lensing power spectrum most notably on large spatial and angular scales, which are 
difficult to access experimentally. Furthermore, cosmic variance and galactic foregrounds prevent accurate measurements on the 
scales in question, i.e. $\gsim\mbox{Gpc}/h$ and above. The small gravitomagnetic corrections could be amplified by cross 
correlation with the kinetic Sunyaev-Zel'dovich effect \citep{1972SZorig}, once future CMB telescopes will provide accurate 
measurements of line-of-sight velocities or with the velocity information from optical galaxy surveys. For contemporary weak lensing surveys, gravitomagnetic corrections to the cosmic shear do not play a significant role.}

\item{The iSW-effect is described by a line-of-sight integration over the divergence of the gravitomagnetic potentials. By this 
argument, the iSW-effect is reduced to a second order lensing effect. Every iSW quantity has a correspondence in weak 
gravitational lensing and the derivation of the power spectrum $C_\tau(\ell)$ proceeds in complete analogy to that of any weak 
lensing quantity, for instance that of the convergence $C_\kappa(\ell)$. The most important difference of the derivation 
presented here to the ones carried out by \citet{1996ApJ...460..549S} or \citet{2002PhRvD..65h3518C} is that our derivation 
explicily pays tribute to the lensing nature of the iSW-effect.}

\item{Gravitomagnetic lensing and the iSW-effect are complementary in measuring the matter flows parallel and 
perpendicular to the line-of-sight. The picture emerging is that (subject to the approximations made) in gravitational light 
deflection (including the gravitomagnetic term $A_\mathrm{z}$), the photon's $\bmath{k}$-vector is rotated but its 
normalisation is conserved. Contrarily, the components of $\bmath{A}$ transverse to the line-of-sight change the normalisation 
of the $\bmath{k}$-vector, i.e. the photon's energy, but leave the direction of $\bmath{k}$ invariant.}

\item{Both effects, gravitomagnetic lensing and the iSW-effect, are achromatic which makes them only accessible by their $n$-point 
statistics. Furthermore, the iSW-effect needs to be separated from other achromatic CMB structures such as the kinetic 
Sunyaev-Zel'dovich effect and the Ostriker-Vishniac effect. The derivation predicts iSW temperature fluctuations of a few $\mu\mathrm{K}$ on large angular scales, which is within reach of future CMB experiments like the 
European {\em Planck}-mission\footnote{\tt http://planck.mpa-garching.mpg.de/}$^{,}$\footnote{\tt 
http://astro.estec.esa.nl/Planck/}, provided that the modelling of Galactic foregrounds is sufficiently accurate to access these large angular scales.}

\item{The gradient $\bmath{\chi}(\bmath{\theta})$ of the iSW temperature fluctuation field $\tau(\bmath{\theta})$ 
should directly map regions of large matter flows, e.g. filaments and clusters with high peculiar velocities, but it can be 
expected to be very susceptible to noise due to the differentiation required in obtaining $\bmath{\chi}(\bmath{\theta})$ from 
$\tau(\bmath{\theta})$, which is reflected by the fact that ratio of the angular power spectra $C_\chi(\ell)/C_\tau(\ell)$ is 
proportional to $\ell(\ell+1)$.}
\end{itemize}

The verification of the theoretical approach by a ray-tracing simulation of photons through a cosmological $n$-body 
simulation will be the subject of a future paper. The non-Gaussian features the iSW-effect and 
gravitational lensing exhibit and the mode-coupling in nonlinear structure growth are unaccessible to perturbation theory and are 
important on small scales. The novel approach to the iSW-effect presented here should allow a much improved 
precision in the numerical treatment, because inaccuracies in interpolating the scalar potential's time derivative 
$\partial\Phi/\partial t$ for each integration time step and in integrating a rapidly oscillating function inherent the direct 
approach \citep[e.g.][]{1995NYASA.759..692T,1995ApJ...445L..73T} are alleviated.

% ############################################################################################## %
% ------------------------------ section: Acknowledgements ------------------------------------- %
\section*{Acknowledgements}
We would like to thank T.~A. En{\ss}lin for reading the manuscript. For numerical evaluation of the Bessel functions $J_\ell(x)$ 
and their derivatives $\dd J_\ell(x)/\dd x$ an exerpt taken from the {\tt CMBfast} code\footnote{\tt http://www.cmbfast.org} 
written by \citet{1996ApJ...469..437S} was used.

% ############################################################################################## %
\appendix

% ############################################################################################## %
% ------------------------------ section: appendix --------------------------------------------- %
\section{Decomposition of mixed 3-point correlators of density and velocity fields}\label{sect_appendix}
In order to evaluate the 3-point correlation function $\bra \delta(\bmath{k}_1)\upsilon(\bmath{k}_2)\delta(\bmath{k}_3)\ket$ in 
perturbation theory, the density- and velocity fields are decomposed into linear terms $\delta^{(1)}, \upsilon^{(1)}$ and 
small perturbations $\delta^{(2)}, \upsilon^{(2)}$:
\begin{equation}
\delta(\bmath{k}) = \delta^{(1)}(\bmath{k}) + \delta^{(2)}(\bmath{k})\mbox{ and }
\upsilon(\bmath{k}) = \upsilon^{(1)}(\bmath{k}) + \upsilon^{(2)}(\bmath{k}).
\end{equation}
As shown by \citet{1984ApJ...277L...5F}, the second order density perturbation can be written as:
\begin{eqnarray}
\delta^{(2)}(\bmath{k}) 
& = & \int\frac{\dd^3 p}{(2\pi)^3}\int\frac{\dd^3 p^\prime}{(2\pi)^3}(2\pi)^3\delta_D(\bmath{p}+\bmath{p}^\prime-\bmath{k})
M(\bmath{p},\bmath{q}) \delta(\bmath{p})\delta(\bmath{p}^\prime) \nonumber \\
& = & \int\frac{\dd^3 p}{(2\pi)^3} M(\bmath{p},\bmath{k}-\bmath{p})\delta^{(1)}(\bmath{p})\delta^{(1)}(\bmath{k}-\bmath{p})
\label{eqn_delta2}
\mbox{,}
\end{eqnarray}
with the function $M(\bmath{p},\bmath{p}^\prime)$ being defined as:
\begin{equation}
M(\bmath{p},\bmath{p}^\prime) = 
\frac{10}{7}+\frac{\bmath{p}\bmath{p}^\prime}{pp^\prime} \left(\frac{p}{p^\prime} + \frac{p^\prime}{p}\right) + 
\frac{4}{7}\left(\frac{\bmath{p}\bmath{p^\prime}}{pp^\prime}\right)^2\mbox{.}
\end{equation}
Clearly, the function $M$ is symmetric, $M(\bmath{p},\bmath{p}^\prime) = M(\bmath{p}^\prime,\bmath{p})$ and has the properties that 
$M(-\bmath{p},-\bmath{p}^\prime)=M(\bmath{p},\bmath{p}^\prime)$ and $M(-\bmath{p},\bmath{p}^\prime)=M(\bmath{p},-\bmath{p}^\prime)$. For the first order perturbation of the velocity field, one obtains:
\begin{equation}
\bmath{\upsilon}^{(2)}(\bmath{k}) = -i H(a) f(\Omega) \frac{\bmath{k}}{k^2}\delta^{(2)}(\bmath{k}).
\end{equation}
The 3-point correlation function $\bra \delta(\bmath{k}_1)\upsilon(\bmath{k}_2)\delta(\bmath{k}_3)\ket$ can now be expanded to 
yield to second order:
\begin{equation}
\bra \delta(\bmath{k}_1)\upsilon(\bmath{k}_2)\delta(\bmath{k}_3)\ket \simeq
\bra \delta^{(1)}(\bmath{k}_1)\upsilon^{(1)}(\bmath{k}_2)\delta^{(2)}(\bmath{k}_3)\ket + (\mathrm{cycl}) +O(2)
\end{equation}
with the zeroth order term $\bra \delta^{(1)}(\bmath{k}_1)\upsilon^{(1)}(\bmath{k}_1)\delta^{(1)}(\bmath{k}_1)\ket$ vanishing due 
to $\upsilon^{(1)}(\bmath{k})\propto\delta^{(1)}(\bmath{k})$ for truly Gaussian random fields. If the perturbation is contained in 
the density field $\delta$, inserting eqn.~(\ref{eqn_delta2}) into the correlator yields:
\begin{equation}
\bra \delta^{(1)}(\bmath{k}_1)\upsilon^{(1)}(\bmath{k}_2)\delta^{(2)}(\bmath{k}_3)\ket =
\int\frac{\dd^3 p}{(2\pi)^3}\int\frac{\dd^3 p^\prime}{(2\pi)^3}\times
\end{equation}
\begin{eqnarray}
\quad (2\pi)^3\delta_D(\bmath{p}+\bmath{p}^\prime-\bmath{k}_3)
M(\bmath{p},\bmath{p}^\prime) \bra\delta(\bmath{k}_1)
\underbrace{\delta(\bmath{p})\ket \bra\delta(\bmath{p}^\prime)}
\upsilon(\bmath{k}_2)\ket
\mbox{.}
\nonumber
\end{eqnarray}
Similarly, if the perturbation is the velocity-field $\upsilon$, one obtains:
\begin{equation}
\bra \delta^{(1)}(\bmath{k}_1)\upsilon^{(2)}(\bmath{k}_2)\delta^{(1)}(\bmath{k}_3)\ket =
\int\frac{\dd^3 p}{(2\pi)^3}\int\frac{\dd^3 p^\prime}{(2\pi)^3}\times
\end{equation}
\begin{eqnarray}
\quad (2\pi)^3\delta_D(\bmath{p}+\bmath{p}^\prime-\bmath{k}_2)
M(\bmath{p},\bmath{p}^\prime) \bra\delta(\bmath{k}_1)
\underbrace{\upsilon(\bmath{p})\ket \bra\upsilon(\bmath{p}^\prime)}
\delta(\bmath{k}_3)\ket
\mbox{.}
\nonumber
\end{eqnarray}
Collecting these results for the mixed 3-point correlator of density and velocity fields in question yields for the first order 
expansion of $\bra \delta(\bmath{k}_1)\upsilon(\bmath{k}_2)\delta(\bmath{k}_3)\ket$ in perturbation theory:
\begin{eqnarray}
\bra \delta^{(1)}(\bmath{k}_1)\upsilon^{(1)}(\bmath{k}_2)\delta^{(2)}(\bmath{k}_3)\ket & = &
M(\bmath{k}_1,\bmath{k}_2) P_{\delta\delta}(\left|\bmath{k}_1\right|) P_{\delta\upsilon}(\left|\bmath{k}_2\right|)
\label{eqn_perturbation_contrib1}\mbox{,}\\
\bra \delta^{(1)}(\bmath{k}_1)\upsilon^{(2)}(\bmath{k}_2)\delta^{(1)}(\bmath{k}_3)\ket & = &
M(\bmath{k}_1,\bmath{k}_3) P_{\delta\upsilon}(\left|\bmath{k}_1\right|) P_{\delta\upsilon}(\left|\bmath{k}_3\right|)
\label{eqn_perturbation_contrib2}\mbox{,}\\
\bra \delta^{(2)}(\bmath{k}_1)\upsilon^{(1)}(\bmath{k}_2)\delta^{(1)}(\bmath{k}_3)\ket & = &
M(\bmath{k}_2,\bmath{k}_3) P_{\delta\delta}(\left|\bmath{k}_2\right|) P_{\delta\upsilon}(\left|\bmath{k}_3\right|)
\label{eqn_perturbation_contrib3}\mbox{,}
\end{eqnarray}
if the condition $\sum_{i=1}^3 \bmath{k}_i=\bmath{0}$ is fulfilled. Hence, in first order perturbation theory, the 3-point 
correlation function can be decomposed into products of the density-density and density-velocity correlation functions, which are 
of the order $\upsilon/c$ (eqns.~\ref{eqn_perturbation_contrib1} and \ref{eqn_perturbation_contrib3}), and into the square of the 
density-velocity cross correlation, which is of order $\upsilon^2/c^2$ (eqn.~\ref{eqn_perturbation_contrib2}).

% ############################################################################################## %
% ------------------------------ section: bibliography ----------------------------------------- %
\bibliography{bibtex/aamnem,bibtex/references}
\bibliographystyle{mn2e}

\bsp

\label{lastpage}

\end{document}